\newif\ifacm
\acmfalse

\newif\ifcomment
\commentfalse

\newif\ifcameraready
\camerareadyfalse

\newif\ifwatermark
\watermarkfalse

\ifwatermark
      \usepackage{draftwatermark}
      \SetWatermarkScale{.7}
      \SetWatermarkText{\shortstack{In Submission:\\Do Not Distribute}}
      \SetWatermarkColor[rgb]{1,0.88,0.88}
\fi

\ifcomment
    \newcounter{MVNumberOfComments}
    \stepcounter{MVNumberOfComments}
    \newcounter{YZNumberOfComments}
    \stepcounter{YZNumberOfComments}
    \newcommand{\mvnote}[1]{\textcolor{blue}{\small \bf [MV\#\arabic{MVNumberOfComments}\stepcounter{MVNumberOfComments}: #1]}}
    \newcommand{\yznote}[1]{\textcolor{red}{\small \bf [YZ\#\arabic{YZNumberOfComments}\stepcounter{YZNumberOfComments}: #1]}}

    \newcommand{\NOTE}[1]
    {
      {\footnotesize\it
        \begin{center}
          \begin{tabular}{|c|}
           \hline
            \parbox{0.85\columnwidth}{
              \medskip
              #1
              \medskip} \\
            \hline
          \end{tabular}
        \end{center}
        }
    }
\else
    \newcommand\mvnote[1]{}
    \newcommand\yznote[1]{}
    \newcommand\NOTE[1]{}
\fi

\newcommand{\eg}{{e.g.,}\xspace}
\newcommand{\ie}{{\it i.e.,}\xspace}
\newcommand{\folder}{./figures}

\newcounter{NumTakeaways}
\stepcounter{NumTakeaways}

\newcommand{\numparticipants}{{31}\xspace}
\newcommand{\numcountries}{{24}\xspace}
\newcommand{\numcities}{{31}\xspace}
\newcommand{\nummobile}{{31}\xspace}

\newcommand{\amigo}{{\texttt{AmiGo}}\xspace}

\newcommand{\testbed}{{\texttt{AmiGo}}\xspace}

\newcommand{\redmi}{{\texttt{Redmi-Go}}\xspace}

\documentclass[10pt, conference, letterpaper]{IEEEtran}

\usepackage{cite}
\usepackage{amsmath,amssymb,amsfonts}
\usepackage{algorithmic}
\usepackage{graphicx}
\usepackage{textcomp}
\usepackage{xcolor}
\def\BibTeX{{\rm B\kern-.05em{\sc i\kern-.025em b}\kern-.08em
    T\kern-.1667em\lower.7ex\hbox{E}\kern-.125emX}}

\usepackage{gensymb}
\usepackage{booktabs}  
\usepackage{pbox}
\usepackage{subfigure}
\usepackage{epsfig,endnotes}
\usepackage{epstopdf}
\usepackage{times}
\usepackage{color}
\usepackage{colortbl} 
\usepackage{graphicx}
\usepackage{xspace}
\usepackage{enumitem}
\usepackage{multirow}
\usepackage{tikz}
\usepackage[ruled,vlined,linesnumbered]{algorithm2e}
\usepackage[T1]{fontenc}
\usepackage[utf8]{inputenc}
\usepackage{url} 

\usepackage{flushend}

\begin{document}
\title{A Worldwide Look Into Mobile Access Networks Through the Eyes of \textit{AmiGos}}
\author{Matteo Varvello$^\dagger$ \hspace{20pt} Yasir Zaki$^\ddagger$ \\
       \hspace{20pt}$^\dagger$Nokia Bell Labs \hspace{20pt}  $^\ddagger$NYU Abu Dhabi
\\
\hspace{20pt}Holmdel, USA \hspace{25pt}  Abu Dhabi, UAE
}
\maketitle

\begin{abstract}
\textit{How does the mobile experience compare between Germany and Nigeria?} There is currently no public data or test-bed to provide an answer to this question. This is because deploying and maintaining such test-bed can be both challenging and expensive. To fill this gap, this paper proposes a novel test-bed design called ``\amigo'', which relies on travelers carrying mobile phones to act as vantage points and collect data on mobile network performance. The \amigo design has three key advantages: it is easy to deploy, has realistic user mobility, and runs on real Android devices. We further developed a suite of measurement tools for  \amigo to perform network measurements, \eg \texttt{pings}, speedtests, and webpage loads. We leverage these tools to measure the performance of \numcountries mobile networks across five continents over a month via an \amigo deployment involving \numparticipants students. We find that 50\% of networks face a 40-70\% chance of providing low data rates, only 20\% achieve low latencies, and networks in Asia, Central/South America, and Africa have significantly higher CDN download times than in Europe. Most news websites load slowly, while YouTube performs well. We made both test-bed and measurement tools open source.
\end{abstract}

\begin{IEEEkeywords}
    mobile, performance, testbed
\end{IEEEkeywords}
\section{Introduction}
\label{sec:intro}
Mobile traffic has been on the rise, even surpassing its desktop counterpart. Data from Google Analytics’ Benchmarking~\cite{analytics} shows that mobile devices drove 61\% of visits to U.S. websites in 2020, up from 57\% in 2019. Desktops were only responsible for 35.7\% of all visits in 2020, and tablets drove the remaining 3.3\% of visitors. Globally, 68.1\% of all website visits in 2020 came from mobile devices.

The research community has investigated how to improve the performance of mobile users, \eg by improving how fast pages load~\cite{speedreader, shandian, polaris, JSCleaner, muzeel, slimweb, jsanalyzer, webtune}, or reducing the data consumption~\cite{flywheel, browselite}. These works are motivated by the \textit{challenging} conditions which mobile users face, such as low bandwidth or high network latencies. However, a comprehensive and open source measurement study of mobile operators in the wild is still lacking, as previous work~\cite{related_hung,ozgu2015monroe} only focused on few mobile operators, mostly in the US and Europe, or few applications. 

Performing such study is challenging due to the lack of a large-scale test-bed in the wild. Even the popular MONROE~\cite{ozgu2015monroe} currently only spans 11 mobile networks in 4 European countries; further, its vantage points are not real mobile devices and have limited mobility. Our first contribution is a novel test-bed design, named \amigo,  which tackles these limitations.  The novelty of the \amigo design lies in the idea to leverage \textit{friends} (hence the name) to carry -- not use -- mobile phones while travelling, guaranteeing network connectivity, \ie WiFi and/or mobile data when possible. We have open sourced all code behind \amigo~\cite{amigoCode} to help measurement researchers deploying their own test-beds in the wild. In the following, we summarize the main contributions of this paper.

\begin{figure*}[thb]
   \centering
   \subfigure[Graphana dashboard.]{\psfig{figure=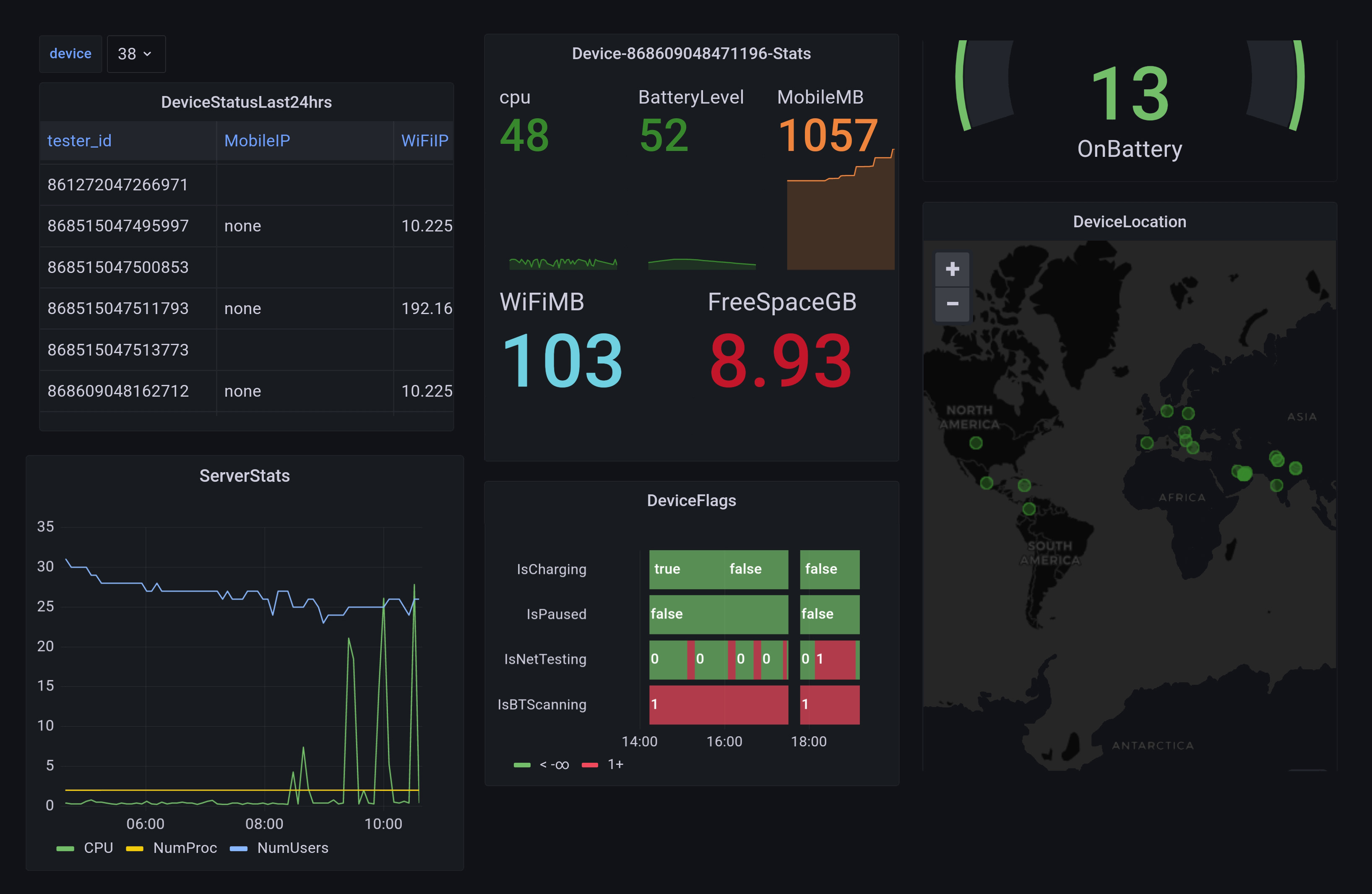, width=2.6in}  \label{fig:amigo_testbed_graphana}}              
   \subfigure[Android application.]{\psfig{figure=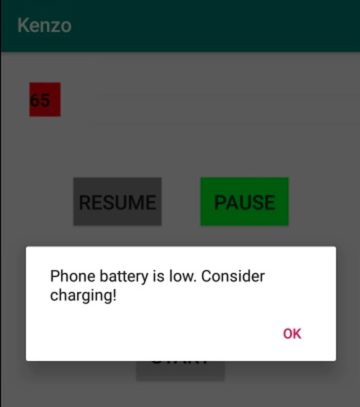, width=1.5in} \label{fig:amigo_testbed_gui}}
   \subfigure[Automated preparation of MEs.]{\psfig{figure=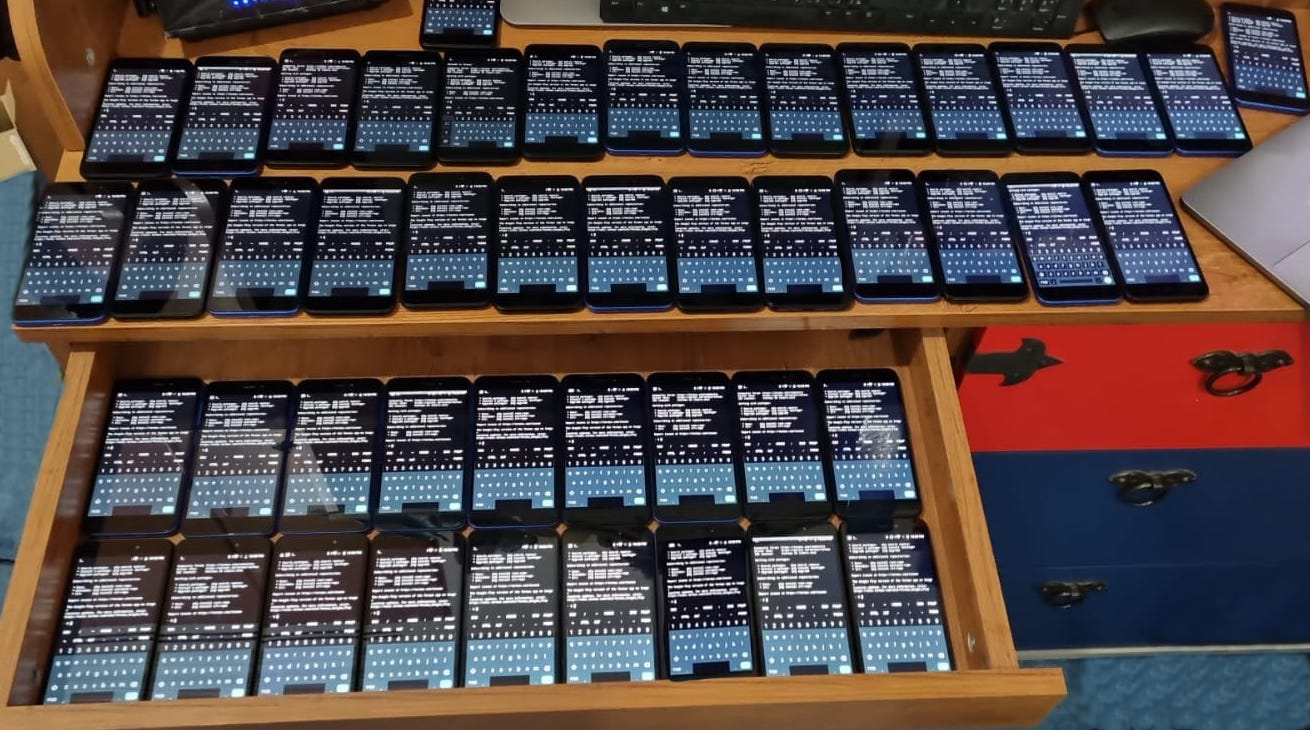, width=2.8in}\label{fig:phones_prepping}}
   \caption{Several components of the \amigo test-bed.}
   \label{fig:amigo_testbed}
\end{figure*}

\vspace{0.05in}
\noindent\textbf{The \amigo test-bed:} It consists of \textit{measurement endpoints}, low-end Android mobile phones (Redmi Go~\cite{redmi}), and a \textit{control server}. These Android devices are rooted and equipped with \texttt{termux}~\cite{termux} -- an Android terminal emulator and Linux environment --  thus allowing fine-grained instrumentation and data collection. The phones frequently report to the control server (running in the cloud) their current status, \eg connectivity (WiFi or mobile) and battery level. The controller monitors the phones' health, schedule experiments, and collect data. Finally, a mobile application allows to interact with the ``amigos'', \eg to pause an experiment or show notifications. 

\vspace{0.05in}
\noindent\textbf{Full stack networking experiments:} We have developed and open sourced~\cite{amigoCode} several tools to perform measurements via \texttt{termux}: speed-tests, traceroutes, DNS queries, HTTP GET of popular CDN objects, webpage loads, and YouTube tests. By combining these tools, we measure multiple metrics \textit{across the stack}, \eg mobile bandwidth, latency to DNS services and content providers, up to user experience when browsing or watching videos. 


\vspace{0.05in}
\noindent\textbf{Measurement campaign:} We recruited \numparticipants university students to participate to (an instance of) the \amigo test-bed  while traveling back to \numcountries different countries during the 2021 winter break (with three countries having more than one student). Overall, this allowed us to collect data across \numcountries mobile networks.  Our key findings are as follows:

\begin{list}{$\bullet$}{\leftmargin=0.5em \itemindent=0em}
    \item \textit{Slow} mobile operators -- download speed wise -- tend to be consistently slow. In contrast, \textit{fast} mobile operators are less consistent. It follows that, even with fast operators, applications that require high bandwidth may suffer from frequent slowdowns. 
    
    \item \textit{Exceptional} latency ($<$20ms) is limited to few mobile networks (less than 20\%) and content providers (Cloudflare and Google). Amazon is most likely to suffer from \textit{less desirable} latencies ($>$150ms) and highly correlated with long network path, suggesting a smaller content distribution network. 
    
    \item Apart from few African operators, DNS resolution on mobile  mostly requires less than 100ms. Several operators (\eg Telenor in Pakistan, IND Airtel in India, and Telcel in Mexico) sometimes rely on Google DNS, we conjecture in presence of network disruptions. 

    \item News websites are often too \textit{heavy} for low-end mobile devices, especially in mobile networks from developing regions. Given the importance of a diversified news outlet, they should prioritize optimizing their website to provide a better user experience for all users, regardless of their device or location. In contrast, YouTube shows better performance across the board. 
    
    
\end{list}






 
\section{Test-bed Architecture}
\label{sec:system}
This section describes the \testbed testbed, which consists of a \emph{control server} to remotely manage mobile \textit{measurement endpoints} (MEs). Our design is generic and can be re-used to build in-house testbeds which aim at controlling multiple devices deployed in the wild. We have thus opened source our code~\cite{amigoCode} to help measurement researchers deploying similar testbeds in the wild.  

\subsection{Control Server}
\label{sec:sys:server}
The control server has three main tasks. First, it monitors the status (battery level, GPS location, etc.) of the MEs. Second, it instruments MEs with \textit{automation instructions} or new commands/actions which are not part of their default behavior, \eg open a reverse SSH tunnel to enable debugging. Third, it maintains a dashboard visualizing the current status of devices and ongoing experiments (see Figure~\ref{fig:amigo_testbed_graphana}).  

The control server is implemented in Python and provides restful APIs which the MEs call to: 1) report their current status (\eg battery level and connectivity), and 2) retrieve \textit{instrumentation code}. The Python code also maintains a \texttt{postgres}~\cite{postgres} database which stores both device status updates and instrumentation code, \ie \texttt{shell} commands to be executed in \texttt{termux} (see below). Finally, \texttt{Graphana}~\cite{graphana} is used to build a visual dashboard allowing to identify potential issues with test-bed and/or experiments.  


\begin{table}[t]
\small
\centering
\begin{tabular}{|l|l|}
    \hline
    {\bf Name} & {\bf Description} \\
    \hline
    vrsNum          & code version number \\
    timestamp       & epoch timestamp at time of report \\
    uid             & unique device identifier  \\
    airplaneMode    & status of airplane mode \\
    googleStatus    & status of Google authorization \\
    uptime          & how long has the device been running \\
    isPaused        & whether the user pause our mobile app \\
    freeSpaceGB     & available space on device \\
    cpuUtilPer      & current CPU utilization \\
    memInfo         & available memory \\
    batteryLevel    & percentage of battery available \\
    isCharging      & whether the device is charging or not  \\
    gpsLoc          & current GPS location \\
    networkLoc      & current network-provided location\\
    foregroundApp   & current app in the foreground  \\
    isNetTesting    & status of network measurements   \\
    wifiIP          & WiFi IP address \\
    wifiSSID        & SSID of WiFi network connected to \\
    wifiQual        & quality of WiFi network signal \\
    todayWiFiData   & data used on WiFi for the day \\
    mobileIP        & mobile IP address \\
    mobileSignal    & quality of mobile network signal \\
    todayMobileData & data used on mobile for the day \\
\hline
\end{tabular}
\vspace{0.1in}
\caption{Summary of data reported by a measurement endpoint to the control server with a 5~min frequency.}
\label{tab:status}
\vspace{-0.2in}
\end{table}

\subsection{Measurement Endpoint (ME)}
\label{sec:sys:endpoint}
\vspace{0.02in}
\noindent\textbf{Rationale and Overview:} Our rationale is to use a real mobile device as a ME. Mobile devices are easy to carry, thus enabling experiments across a plethora of networks and realistic conditions. They support multiple access networks, which allows flexibility for both remote instrumentation (\eg via WiFi) and experiments (\eg WiFi or mobile, even switching between 3G and 4G). Mobile devices also support a wide range of experiments, \eg from \texttt{ping} to Web browsing. Finally, the data collected via a real device is representative of what experienced by actual users in the wild. 

We chose a \redmi~\cite{redmi} as \amigo's ME. This device is based on Android which allows full instrumentation while being the most popular operating system in the world~\cite{android_stats}. It is also a cheap device (retail price of \$70), which is paramount when the goal is to realize a test-bed with a large number of MEs. Note that despite being a low-end device (1.4~GHz quad-core CPU and 1~GB of RAM) our benchmark shows little to no impact to most measurements (with the exception of webpage loads, see Section~\ref{sec:res:perf}). Third, it can be easily rooted, allowing full control via \texttt{termux}~\cite{termux}, an Android terminal emulator and Linux environment. This gives us the flexibility to write \textit{any} instrumentation, from low level utilities like \texttt{ping} to app automation, and data collection, even when root privilege is needed. We later discuss how this extends to other devices. 

\vspace{0.05in}
\noindent\textbf{Design and Implementation:} We refer to the core code of the ME as \amigo's \textit{agent}. The agent has overall three key tasks. First, it monitors the endpoint resources, \eg battery level and connectivity. Second, it interacts with the control server, either to report its state (see Table~\ref{tab:status}) or to collect instrumentation for new experiments to run. We leverage a combination of classic Linux tool (\eg \texttt{ifconfig}), Android tool (\eg \texttt{dumpsys}), and termux tools (\eg \texttt{termux-location}) to populate the state information which is reported to the control server every 5 minutes. Third, it controls when to run a list of pre-installed experiments by matching a set of rules, \eg only run with a given frequency when on mobile. Each experiment consists of a \texttt{shell} script, thus allowing for easy customization of the ME's behavior (see Section~\ref{sec:data:meth}).

We use \texttt{cron} to ensure the agent is started at each phone reboot, \eg in case the device running out of battery. Further, we use it to force a device reboot each night to ``clean'' potential wrong states reached during any of the experiments. We use the \texttt{Boot Apps} Android app~\cite{boot_apps} to ensure \texttt{termux} is launched at reboot, which in turns enable \texttt{cron} and \texttt{sshd}. 


Each device is instrumented with a private SSH key used to: 1) pull code updates from \texttt{github}~\cite{github}, 2) open a reverse tunnel to the control server, if requested to do so, which is useful for debugging misbehaving MEs. To guarantee a consistent state across all devices -- and speedup the test-bed preparion job -- the installation of every app and termux package is automated and relies on local packages provided by APKMirror~\cite{apkmirror}. Figure~\ref{fig:phones_prepping} shows the automated preparation of multiple MEs. 

\vspace{0.05in}
\noindent\textbf{Mobile Application:} It acts as \amigo's GUI and accomplishes three tasks. First, it shows the device identifier for simple communication between test-bed maintainers and a volunteer carrying the ME in presence of a concern. Second, it allows to pause an ongoing experiment. This is required when the volunteer  needs to interact with the device, \eg to connect to a WiFi network. Without this feature, any ongoing experiment could collide with the user causing potential issues. Finally, it notifies the user when the mobile phone requires charging (see Figure~\ref{fig:amigo_testbed_gui}), which is often neglected by volunteers (see Section~\ref{sec:conclusion}). 

\vspace{0.05in}
\noindent\textbf{Google Account:} Android devices require a Google account. We have contacted Google asking for a testing account, with no luck. We have thus setup each device with the same Google account, given that no limitation on the number of devices exists. This has triggered random requests to verify our account, by entering its username and password. Fortunately, this operation can be automated by detecting the presence of the \texttt{MinuteMaidActivity} when launching YouTube. 

\vspace{0.05in}
\noindent\textbf{Limitations:} The main limitation of the \amigo agent is that it currently targets a specific Android phone: Redmi Go. We purposely focused on one device to enable fair comparison of results across different mobile operators. Still, it would be interesting to extend to more recent and powerful devices, \eg to investigate 5G connectivity. Extending to other Android devices is simple -- as long as they can be rooted -- as \amigo's agent only relies on low level instructions shared among Android devices. Extending to iOS would also be of interest although hard due to its limited automation capabilites. 

\begin{figure}[t]
    \centering
    \subfigure[CPU utilization per test.]{\includegraphics[width=0.5\linewidth]{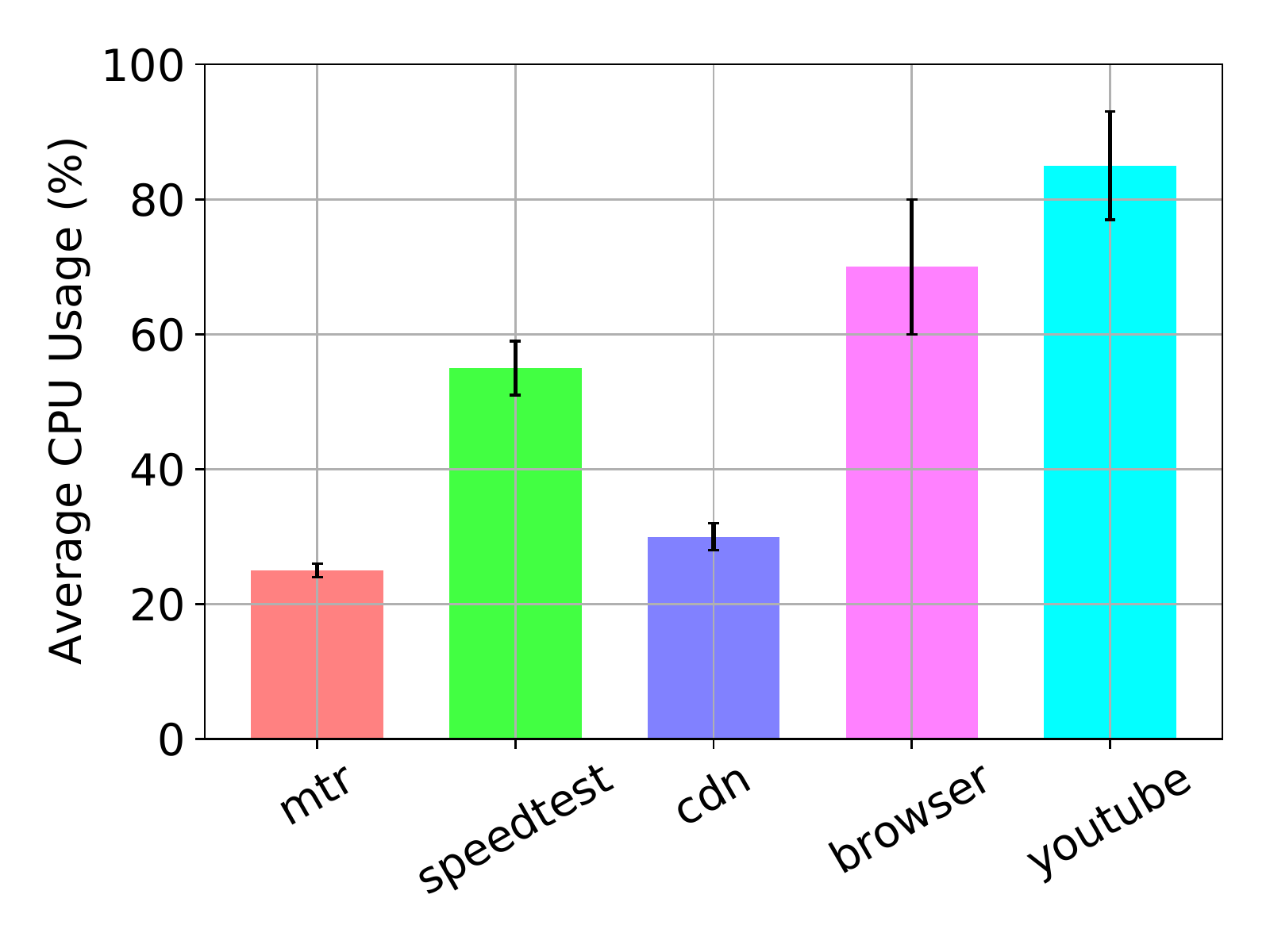}\label{fig:cpu}}\hfill 
     \subfigure[Memory utilization per test.]{\includegraphics[width=0.5\linewidth]{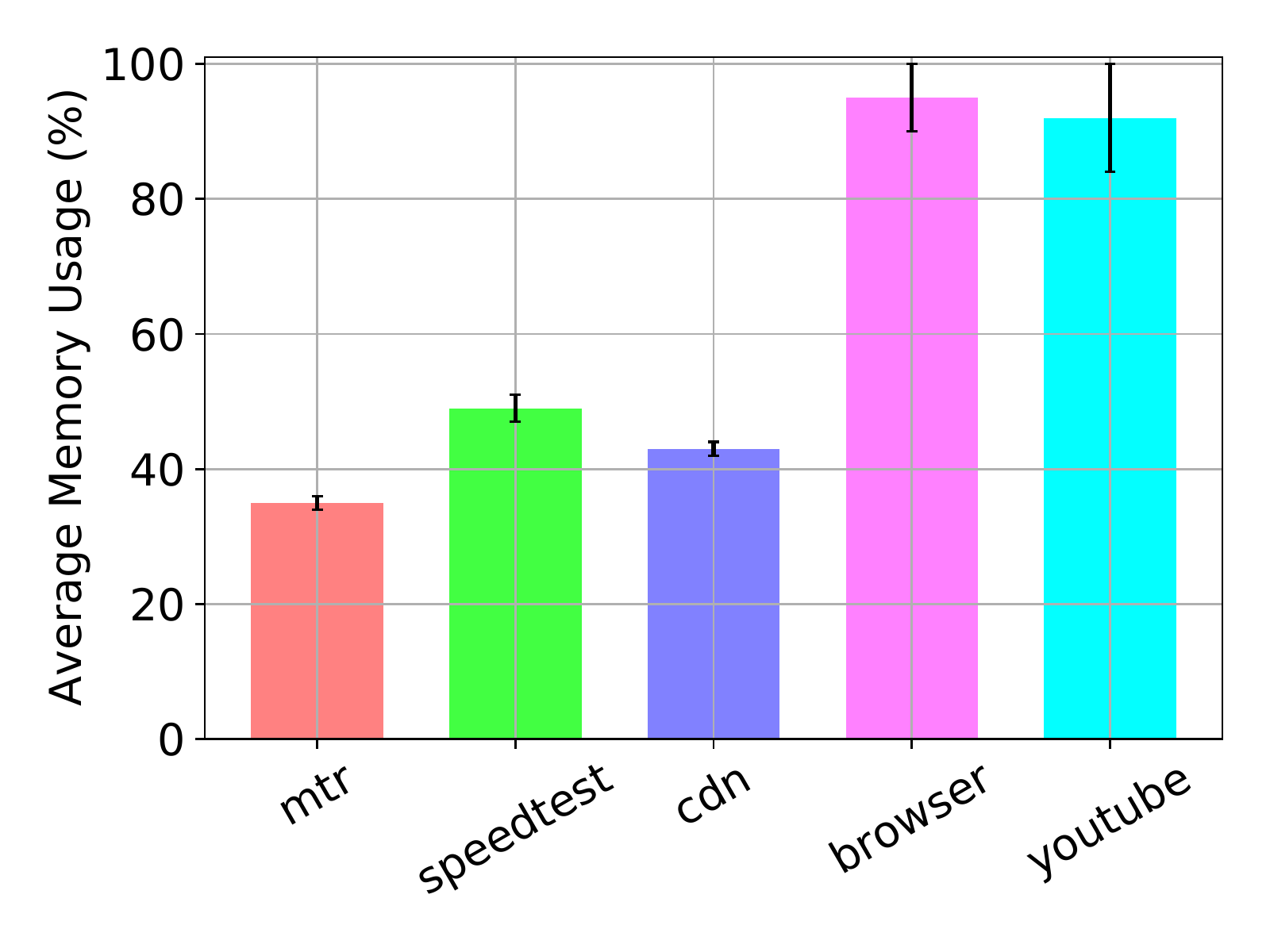}\label{fig:mem}}\hfill 
    \caption{Benchmarking of CPU and memory usage at the ME (\redmi with 1.4~GHz quad-core CPU and 1~GB of RAM). Each bar reports the average CPU/memory usage per networking test. Errorbars report standard deviation.}
    \label{fig:bench}
\end{figure}

A potential additional issue with the device selection is that it might have an impact on the measurements performed, due to its limited hardware (1.4~GHz quad-core CPU and 1~GB of RAM). We have benchmarked CPU (Figure~\ref{fig:cpu}) and memory (Figure~\ref{fig:mem}) usage when performing the measurements described in Section~\ref{sec:data:meth}. CPU-wise, most experiments are not concerning given that the device's CPU is rarely under stress, only in some cases YouTube approaches 90\% CPU usage. The same is not true for memory usage which is instead fully occupied for both browser and YouTube experiments. This implies that higher level experiments can be partially impacted by the device we chose. We acknowledge this as a potential limitation due to the need to keep the cost per device low, and reach a large number of mobile networks.

\begin{figure*}[!hbt]
    \subfigure[\texttt{Germany}]{\includegraphics[width=0.24\linewidth]{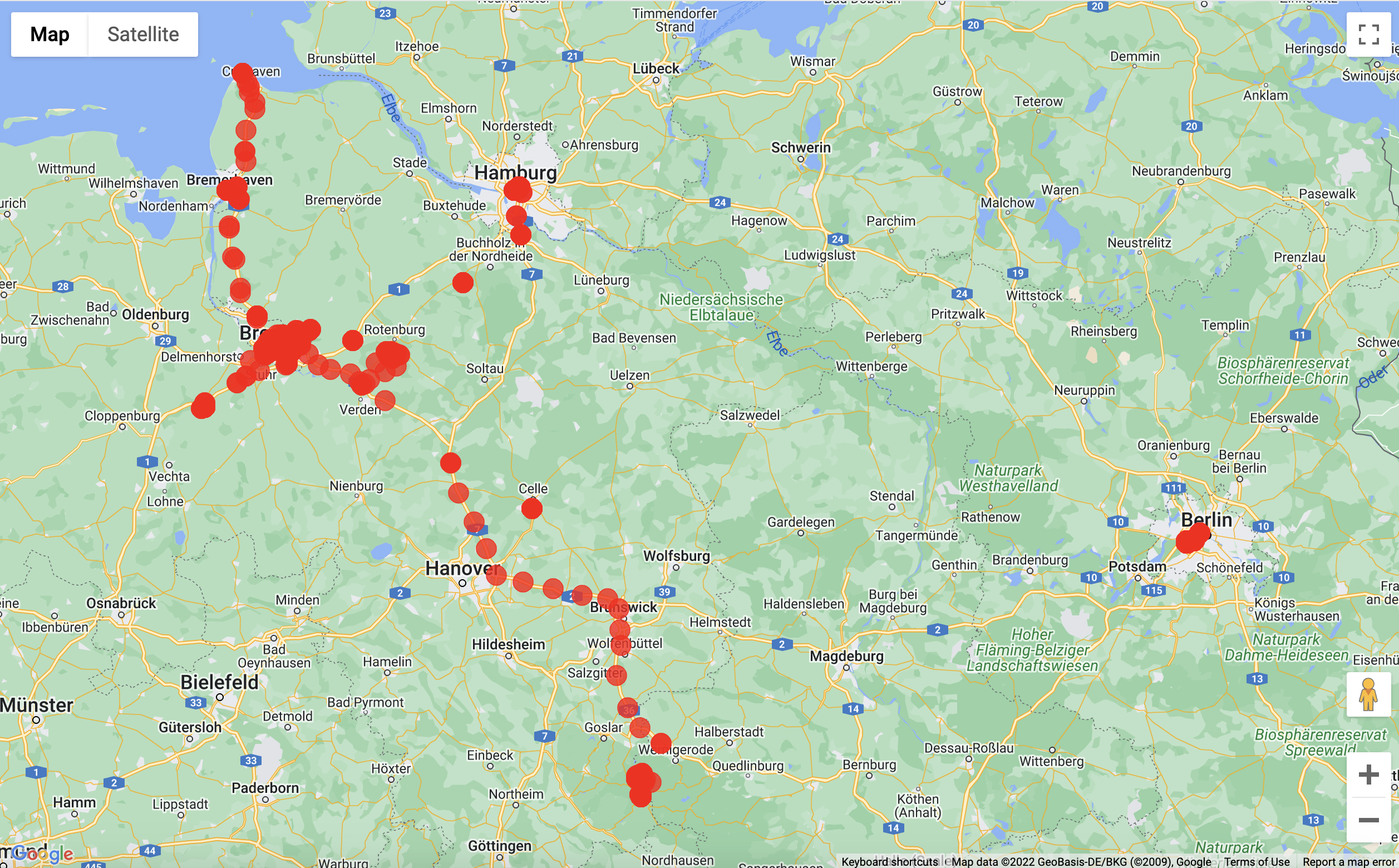}}\hfill
    \subfigure[\texttt{Jamaica}]{\includegraphics[width=0.24\linewidth]{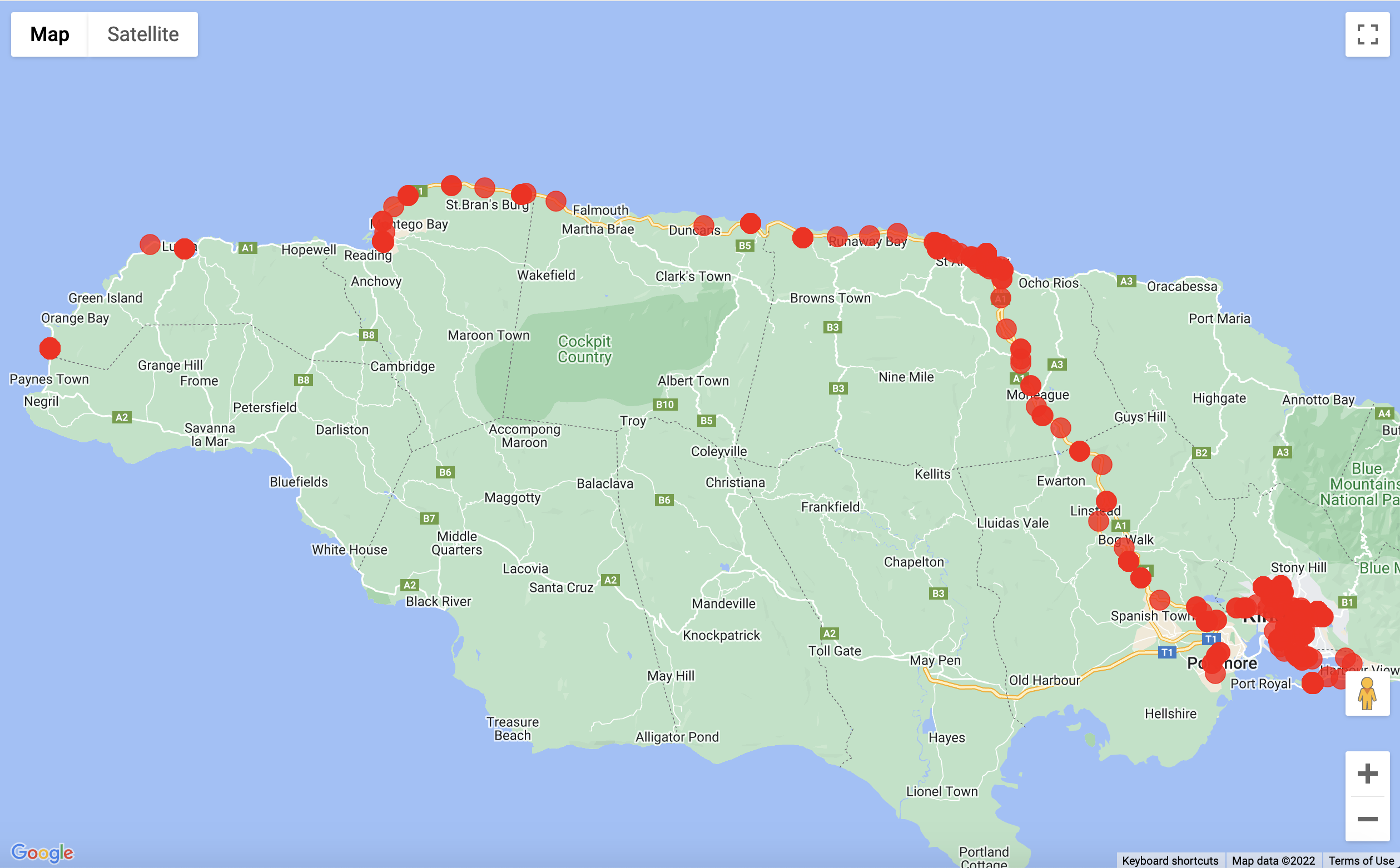}}\hfill
    \subfigure[\texttt{Tunisia}]{\includegraphics[width=0.24\linewidth]{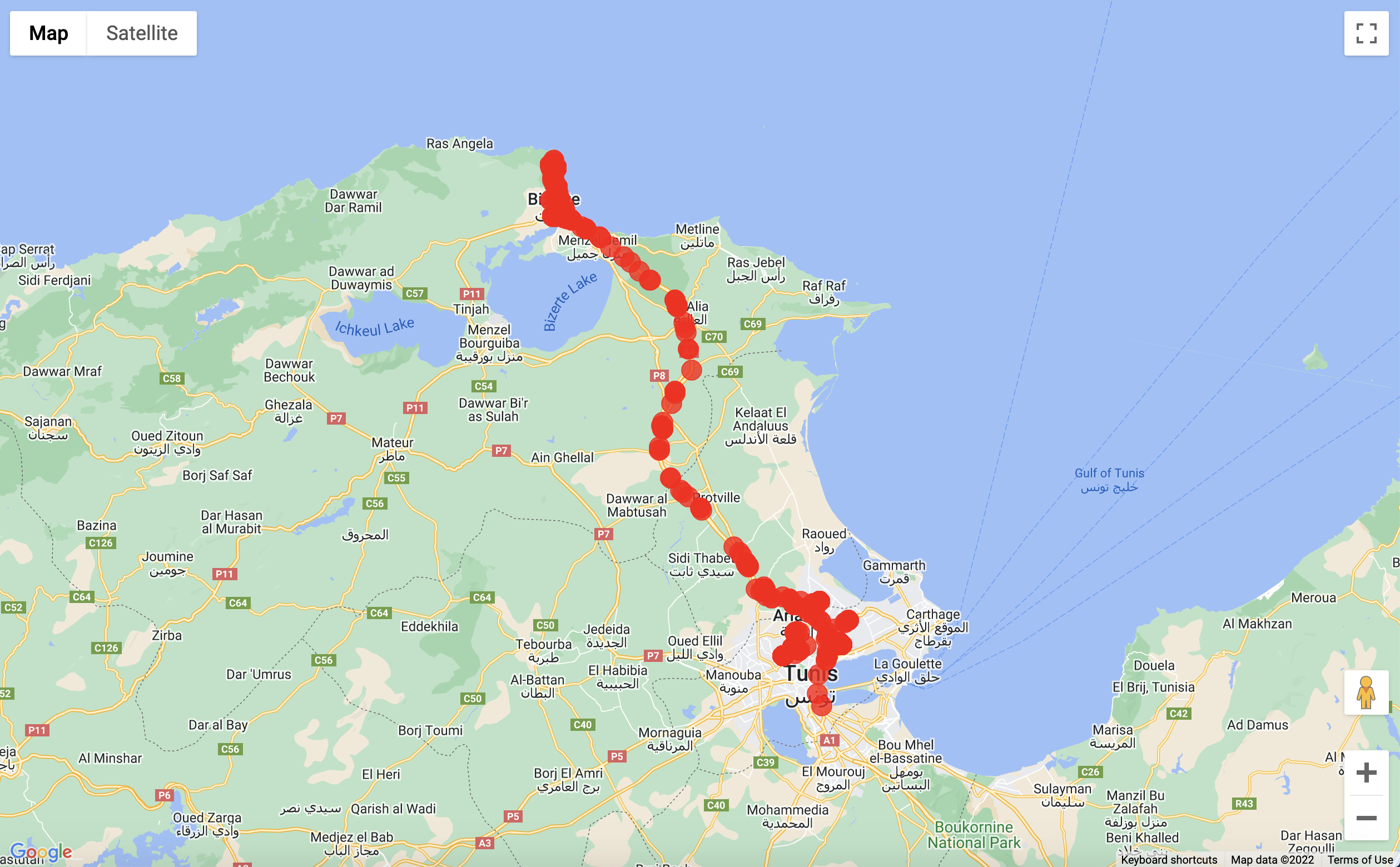}}\hfill
    \subfigure[\texttt{Portugal}]{\includegraphics[width=0.24\linewidth]{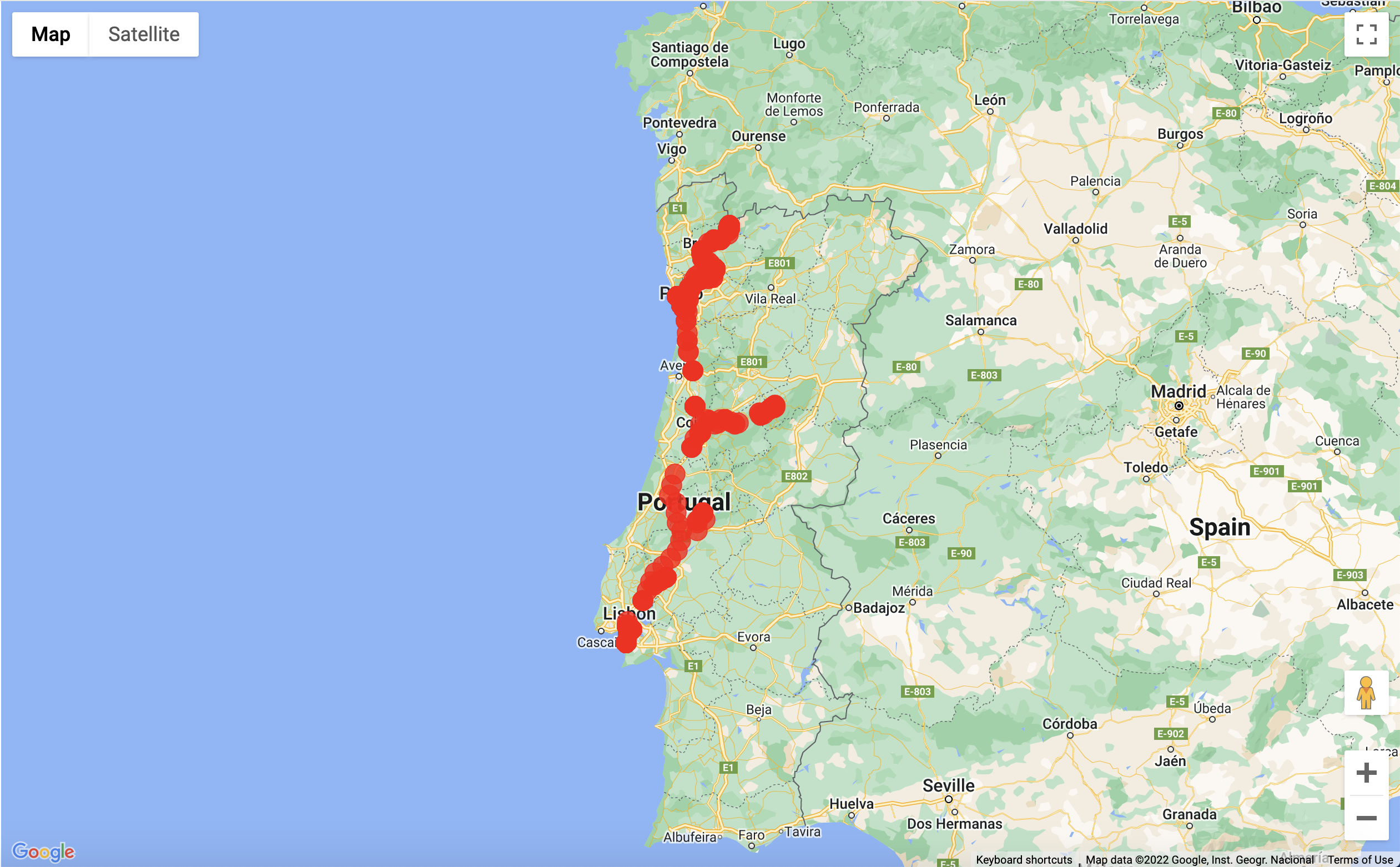}}\hfill 
    \caption{Visualization of \amigo volunteers mobility in four countries: Germany, Jamaica, Tunisia, and Portugal.}  \label{fig:amigo:maps}
\end{figure*}

\begin{figure}[t]
    \center
    \includegraphics[width = \linewidth]{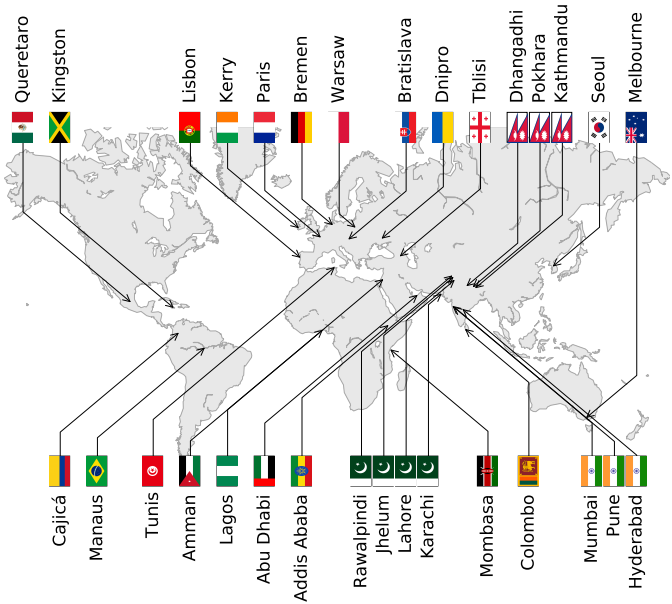}
    \caption{Geographical distribution of the \amigo test-bed deployment between December 2021 and January 2022.}
    \label{fig:locations}
\end{figure}

\section{Data Collection}
\label{sec:data2}
This section describes the experiments we devised to measure the performance of mobile networks using the \amigo test-bed. We instrument the agent to schedule experiments every 30 minutes when the ME is \textit{only} connected to a mobile network. Note that we asked volunteers to connect the ME to their home WiFi as a simple mean to avoid running most experiments from their home, where we expect them to spend the majority of the time. We further monitor data consumption to avoid consuming more than 4~GB per day. As for the test-bed code, we have also open sourced it for each experiment. 

\subsection{Experiments Description}
\label{sec:data:meth}

\vspace{0.05in}
\noindent\textbf{Access Characteristics}: Mobile networks are characterized by three main metrics: download/upload speed, latency, and loss probability. We use a combination of tools to compute each metric. First, we measure download/upload speeds using \textit{Speedtest CLI}~\cite{speedtest_cli}, a Linux-native Speedtest tool backed by Ookla$^{\circledR}$.
Next, we derive network losses from pcap files captured via \texttt{tcpdump} while loading popular webpages. Finally, we use \texttt{mtr}~\cite{mtr} -- a network diagnostic tool combining ping and traceroute -- to derive latency and network path towards popular content providers (\texttt{Amazon}, \texttt{Facebook}, \texttt{Google}, \texttt{YouTube}) and DNS providers (\texttt{Google} and \texttt{Cloudflare}). The rationale, as suggested in~\cite{10.1145/3487552.3487815, 10.1145/3487552.3487854}, is that these providers employ edge nodes which are commonly close to the users. 

\vspace{0.05in}
\noindent\textbf{DNS Performance}: DNS is a critical component of every mobile application. We instrument our MEs to use the DNS provided by the mobile network they are connected to.  We evaluate DNS on mobile using statistics extracted from pcap files of webpage loads. 

    
\vspace{0.05in}
\noindent\textbf{CDN Performance}: A Content Delivery Network (CDN) is a popular networking tool used to accelerate content retrieval, thus improving the user experience, by moving popular content close to a user's network location. We have identified a popular JS file (\texttt{jquery.min.js} version 3.6.0, or the last version at the time of our measurement campaign) which is hosted at multiple CDNs (\texttt{Cloudflare}, \texttt{Facebook CDN}, \texttt{Google CDN}, \texttt{Highwinds CDN}, \texttt{jsDelivr}, and \texttt{Microsoft Ajax CDN}). Note that \texttt{jsDelivr} advertises optimal performance by matching each request to an ``optimal'' CDN based on uptime and performance~\cite{jsdelivr}. 


We iterate through these CDNs fetching \texttt{jquery.min.js} using \texttt{cURL} instrumented to report the total download time. We further collect HTTP headers since, for some CDNs, they report whether the file was found in a cache, or not. Specifically,  \texttt{Cloudflare}, \texttt{jsDelivr}, and \texttt{Microsoft Ajax} report cache $HIT$s or $MISS$es using two HTTP response header fields: \texttt{x-cache} and \texttt{cf-cache-status} (from \texttt{Cloudflare}~\cite{cloudflare}).  \texttt{Fastly} also use the \texttt{x-cache} header to report (\eg `$HIT, MISS$', `$MISS, HIT$') \textit{where} a cache hit or miss has occurred: either at the edge (second entry) or at the  \texttt{shield} (first entry). A shield is a mid-tier caching layer between origin server(s) and edge servers~\cite{Fastly}.  Since \texttt{jsDelivr} relies on a network of CDNs, including \texttt{Fastly}, in some cases it also reports where a cache hit/miss has occurred. 


\begin{figure*}[!tb]
    \centering
    \includegraphics[width=\linewidth]{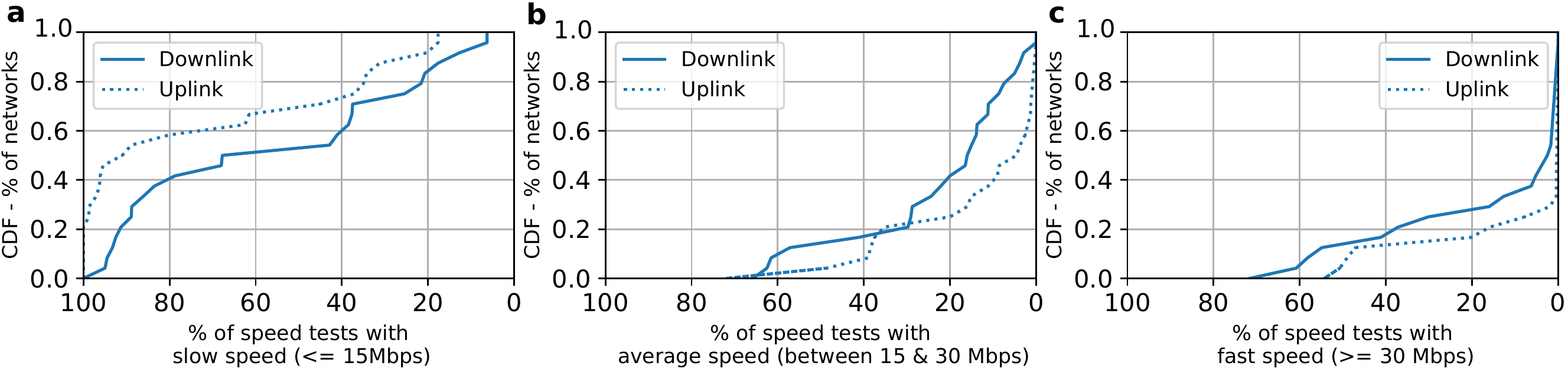}
    \caption{Speed tests' measurements results showing the CDF of percentage of mobile networks with downlink and uplink speeds quantified as: (a) slow ($<=$15Mbps), (b) average (between 15 and 30 Mbps), (c) fast ($>=$30Mbps).}
    \label{fig:speedtest_aggregate}
\end{figure*}

\vspace{0.05in}
\noindent\textbf{Application Performance}: Web browsing and video streaming are two  popular mobile applications which are \textit{easy} to automate, differently from even more popular apps like TikTok or Instagram. For Web measurements, we load several news websites via Google Chrome, namely: \texttt{cnn.com}, \texttt{wsj.com},  \texttt{bbc.com}, \texttt{foxnews.com}, and \texttt{washingtonpost.com}. During a page load we record \texttt{pcap} traces and a video which is then fed to \texttt{visualmetrics}~\cite{visualmetrics} to extract performance timing metrics such as SpeedIndex~\cite{lighthouse} -- which reports the time it takes for the visible parts of a webpage to be displayed.

For video streaming, we automate YouTube. This automation requires interacting with its GUI and is thus specific to the Redmi Go, or better the version of YouTube under test (17.43.46) and a device with a screen resolution of 720x1280 pixels. Extending to other devices is straightforward, given the logic of the automation does not change, but it requires some manual verification. YouTube allows to enable ``stats-for-nerds''~\cite{stats_nerds} which report information like buffer occupancy and number of lost frames. Such information is reported on screen, over the video, and can then be copied on the clipboard and dumped to a file. 
We also collect \texttt{pcap} traces while performing YouTube experiments. We used a (up to) 4K video specifically produced for testing (\textit{https://www.youtube.com/watch?v=TSZxxqHoLzE}).




\subsection{Data Overview}
\label{sec:data:overview}
\vspace{0.05in}
\noindent\textbf{World Coverage:}  We have deployed \amigo MEs via \numparticipants university students while travelling back home during the 2021 winter break. The students were recruited based on the home countries  they were traveling back to, as well as their willingness to carry the phones as much as possible. The data collected spans \numcities cities in \numcountries countries (see Figure~\ref{fig:locations}) between December 15th, 2021 and January 30th, 2022. Multiple students travelled to different cities in three countries: 4 to Pakistan, 3 to India, and 3 to Nepal. The students were instructed to purchase a mobile SIM card from their destination with a data plan around 40/50~GB. The choice of the mobile network operator was left to the students based on their own convenience. The students did move within each country, sometimes even going to multiple cities. Figure~\ref{fig:amigo:maps} shows  sample users' movements in four different countries. 

\vspace{0.05in}
\noindent\textbf{Mobile Access Type:}  Overall, the vast majority of the measurements were performed on 4G: 60\% of the phones had 90\% of their measurements performed on 4G. Nevertheless, 10\% of the phones never encountered a 4G mobile network during their tests. \mvnote{@yasir - what we do with this 10\%? is it included in the data?}\yznote{If I remember correctly the answer is yes. We include them.} \mvnote{Don't you think we can be attacked for this? aka, should we say we filter 3G?}\yznote{But why would we be attached. If some mobile networks are bad and are switching to 3G that is a performance degradation which should be part of the analysis (I think?).} The remainder 30\% of phones had 40\% - 90\% of their tests performed on 4G.  

\vspace{0.05in}
\noindent\textbf{User Mobility:} We use GPS data to monitor user mobility. Most users were quite mobile, exploring more than 200~km during the measurement campaign and spending about 100~hrs outdoor (roughly 4.5 days). Some users ventured in trips; for example, one user went from France to Spain, resulting in being located 700~km away from what was marked as the home location. In few cases, users spent very little time outside due to safety restrictions caused by the rapid spread of Omicron (COVID-19 variant)~\cite{omicron}.



\vspace{0.05in}
\noindent\textbf{Data Usage:}  Data consumption depends on two factors: 1) how much users move, 2) the 4~GB limit per day we set. The rationale of the latter limit was not to consume more than 40/50~GB (the target data plan we recommended acquiring) over a 10/15 days period (the average expected travel duration), so to avoid the inconvenience to recharge the data plan. Overall, our daily limit was very conservative since the median total data consumption was close to 6~GB. 
Few participants (Poland and Germany) consumed a lot of mobile data: 44.9~GB (over 44 days) and 26~GB (over 29 days), respectively. This was driven by high user activity and increased trip duration due to COVID-related travel restrictions.

\subsection{Ethics}

Given that we recruited participants to carry instrumented mobile devices (\amigo MEs), we obtained an institutional review board (IRB) approval (HRPP-2021-185) to conduct these studies. In addition, one of the authors have completed the required research ethics and compliance training, and was CITI ~\cite{citi} certified. Participants were also provided with a consent form to read, and sign, acknowledging their willingness to participate. Further, they were given the opportunity to ask questions about the study and what data was being collected. 

We asked participants to carry \amigo MEs, charge them, install a mobile data sim-card, and connect them to WiFi when possible. We also instructed the participants not to use the mobile phones or add any of their personal information or logins. As such, we do not collect any unidentifiable, sensitive, or personal information about the participants. The only foreseeable concern that might put our users' privacy at risk is the collection of the phones' GPS data, which in principle can reveal the participants movements. We informed the participants beforehand about this concern and we obtained their written consent that they approve this collection. 



\vspace{0.2in}
\section{Network Characterizations}
\label{sec:res:network}
This section analyzes low-level networking experiments, \eg speedtests,  traceroutes, and DNS lookups. For a given metric, our data-set contains multiple data points for each mobile operator. Instead of plotting one Cumulative Distribution Function (CDF) per mobile operator, to ease visibility we plot the CDF of the percentage of mobile networks for which a target metric is within a range. This analysis, inspired by Chrome User Experience Report (CRuX)~\cite{crux-data}, allows to comment on the probability, expressed in number of tests, that a certain condition is met at a fraction of the mobile networks. 



\vspace{0.1in}
\noindent\textbf{Speed Tests}: Figure~\ref{fig:speedtest_aggregate} shows the CDF of the percentage of mobile networks that have \textit{slow}, \textit{average}, and \textit{fast} downlinks (solid) or uplinks (dashed). A slow downlink/uplink is characterized by a speed smaller than 15~Mbps; average refers to a speed between 15 and 30 Mbps. Finally, fast refers to a speed higher than 30~Mbps. These thresholds are drived form the SpeedTest Global Index which ranks 140 operators by their upload/download speeds~\cite{speedtest_data}. 

As also reported in~\cite{speedtest_data}, Figure~\ref{fig:speedtest_aggregate} shows that slow speeds are more likely on uplink than on downlink, \eg 90\% of the uploads were slow for 50\% of the mobile networks (whereas such frequent slow downloads only happen for 25\% of the  mobile networks). Average and fast speeds are instead more likely in download than upload. For both upload and download speeds, slow tests are more \textit{consistent}, meaning that a slow mobile operator tends to be slow for the majority of the test, \eg 40\% of the operators have constant speed lower than 15~Mbps for 80-100\% of the tests (\eg Flow in Jamaica). Conversely, fast mobile operators are less consistent, \ie most have less than 50\% of the measurements which can be considered fast (\eg Telcel in Mexico). This result suggests that applications which constantly require high bandwidth, \eg volumetric videos or virtual/augmented reality, might suffer from frequent slowdowns. 

\begin{figure}[!tb]
    \centering
    \includegraphics[width=\linewidth]{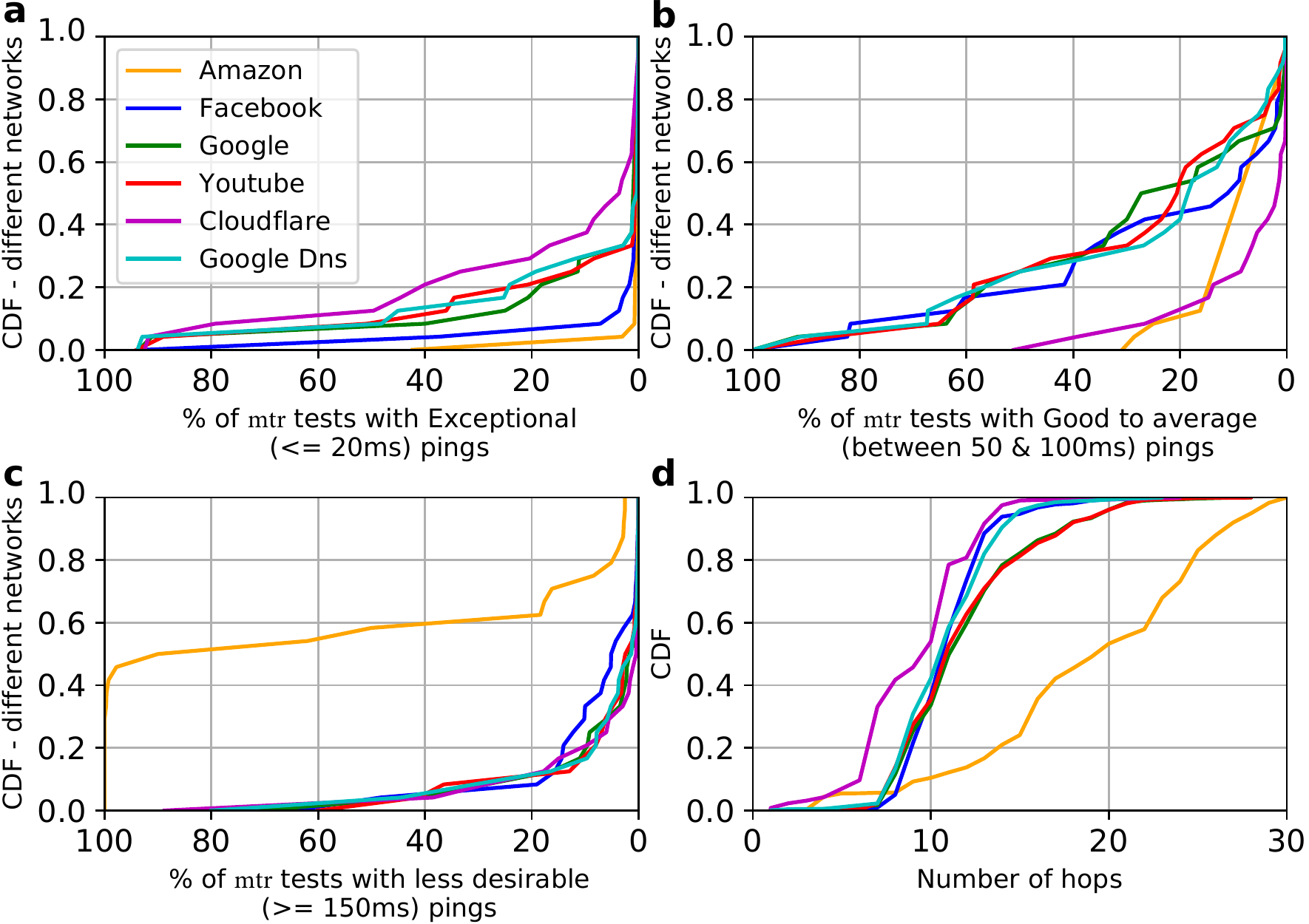}
    \caption{CDFs of percentage of mobile networks with latencies per content provider classified as: (a) exceptional ($\le20ms$), (b) good to average (between $50ms$ and $100ms$), (c) less desirable ($\ge150ms$). Figure (d) shows CDFs of the number of hops between MEs and each content provider. Latency and number of hops results are obtained using \texttt{mtr}.}
    \label{fig:MTR_aggregate_ping}
\end{figure}


\vspace{0.1in}
\noindent\textbf{Mtr Pings and Traceroute}:  Figure~\ref{fig:MTR_aggregate_ping} (a), (b), and (c) show respectively the aggregated results based on the quality of the mobile network latency tests~\cite{ping_report}: 1) \textit{exceptional} ping which falls below 20~ms, 2) \textit{good} to \textit{average} ping between 50 and 100~ms, and 3) \textit{less desirable} ping which exceeds 150~ms. These figures also show the latency CDFs split based on the tested domain: \texttt{Amazon}, \texttt{Facebook}, \texttt{Google}, \texttt{YouTube}, \texttt{Cloudflare DNS}, and \texttt{Google DNS}. 

Figure~\ref{fig:MTR_aggregate_ping} (a) shows that exceptional latencies are rare. Only a small percentage of mobile networks (\ie 20\%) have exceptional latencies to about four of the tested servers: \texttt{Cloudflare} (40\% of tests), and \texttt{Google} products (DNS, YouTube, and search engine) with 20-25\% of the tests. Conversely, both \texttt{Facebook} and \texttt{Amazon} have nearly no tests with exceptional latencies. For the ``good to average'' pings, apart from \texttt{Cloudflare} and \texttt{Amazon},  Figure~\ref{fig:MTR_aggregate_ping} (b) shows that latency to all other popular providers are similarly split between mobile networks and number of tests. For example, 20\% of mobile networks have about 60\% of their tests ranked as ``good to average''; this suggests that such ping values are mostly due to network conditions at the mobile networks rather than provider performance. \texttt{Amazon} and \texttt{Cloudflare} are outliers due to their limited number of samples in this category: indeed, the majority of their samples fall under exceptional pings (in the case of \texttt{Cloudflare}) and less desirable (in the case of \texttt{Amazon}), for which 50\% of mobile networks have 80\% less desirable pings. To better understand the previous results, we investigate the length of the network path between mobile users and each measured service. Figure~\ref{fig:MTR_aggregate_ping} (d) shows that \texttt{Cloudflare} and \texttt{Amazon}  have the shortest and longest number of hops to our mobile users, thus confirming their latency results observed above.





\begin{figure}[!tb]
    \centering
    \includegraphics[width=\linewidth]{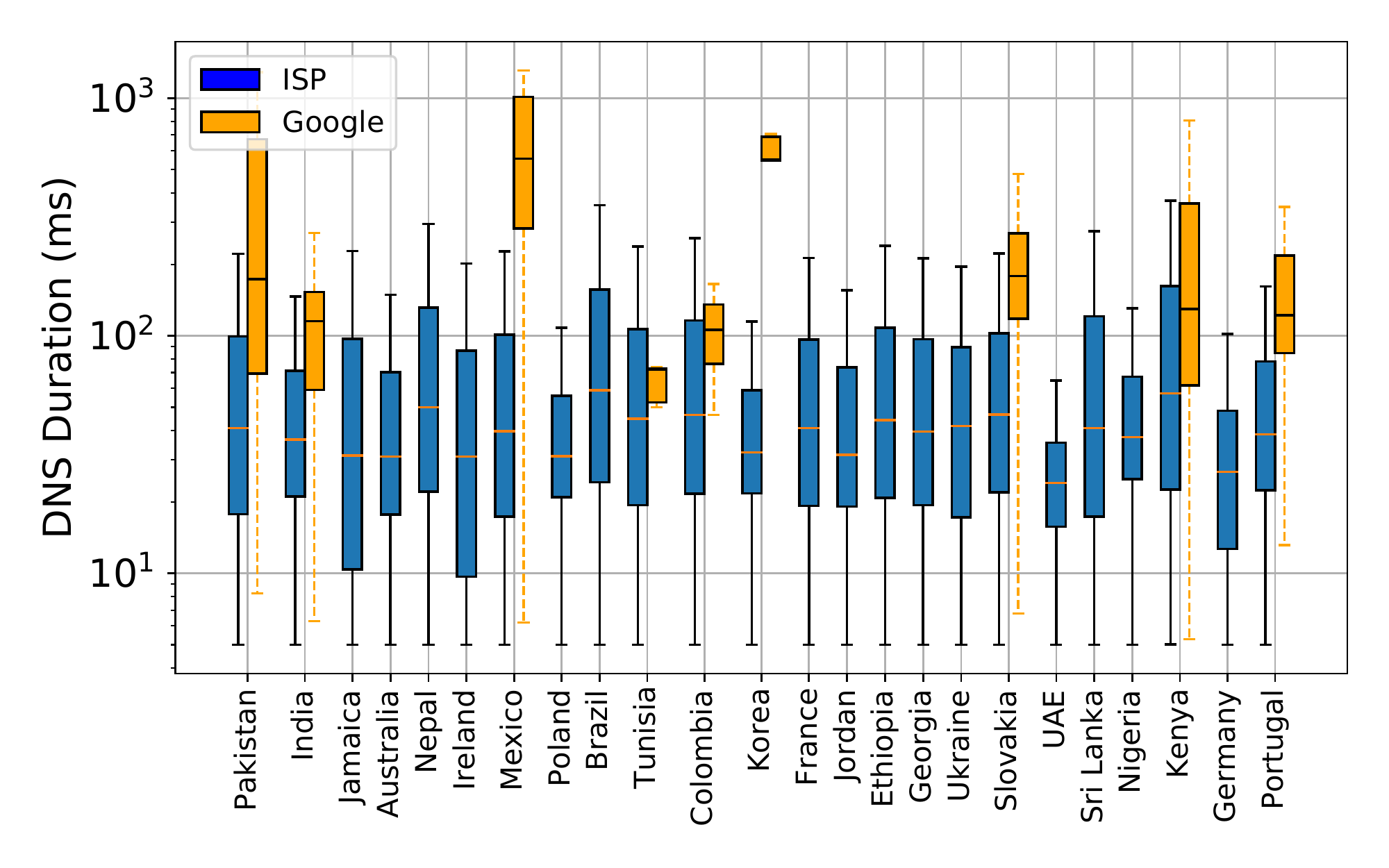}
    \caption{Box plots of DNS lookup times per mobile operator, comparing when the phones used local DNS servers (provided by the mobile operator) versus Google DNS, if available. Note that Google DNS was not setup on the phone, but likely used by the mobile operator in presence of  outages or high load.}
    \label{fig:dns}
    \vspace{-0.15in}
\end{figure}

\begin{figure*}[t]
    \centering
    \subfigure[CDN download time as a function of cache HIT or MISS.]{\includegraphics[width=0.32\linewidth]{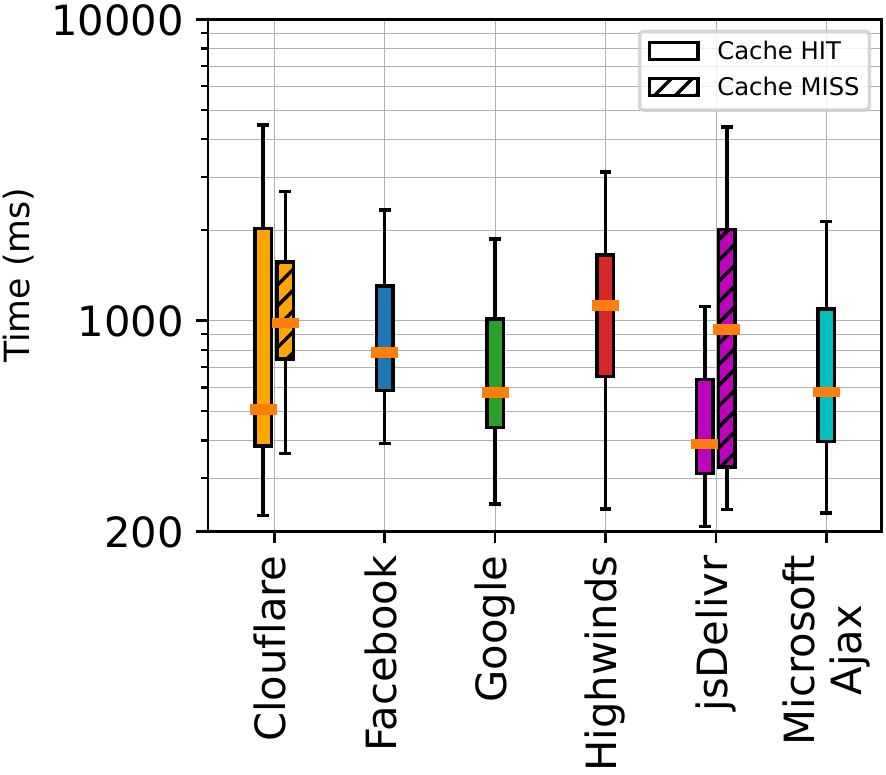}\label{fig:CDN_aggregate_hit_miss}}\hfill   
    \subfigure[CDN download time per continent.]{\includegraphics[width=0.64\linewidth]{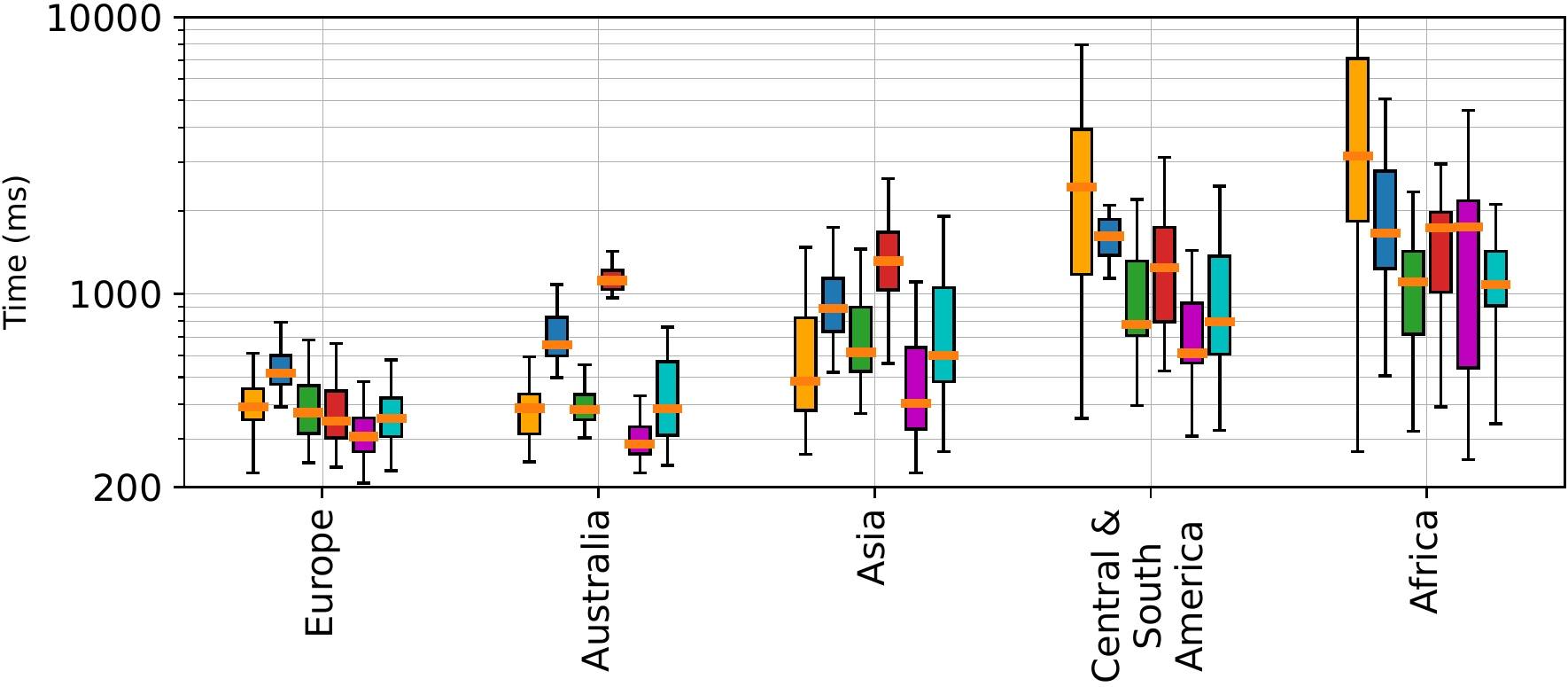}\label{fig:CDN_aggregate_continent}}\hfill 
    \caption{Box plots of total download time for six popular CDN servers: (a) overall comparison of the download time for each CDN, including the impact of cache MISSes (for \texttt{Cloudflare} and \texttt{jsDelivr}), (b) comparison of the download time split per five continent for each CDN server.}\label{fig:CDN_aggregate}
\end{figure*}

\vspace{0.1in}
\noindent\textbf{Domain Name System}: Figure~\ref{fig:dns} shows box plots of DNS lookup times---extracted from Web browsing tests--- per mobile operator. Blue box plots refer to lookup times obtained when a \textit{local} (provided by the mobile operator) DNS is used; orange box plots refer to lookup times obtained via Google DNS---identified by \texttt{8.8.8.8} or \texttt{8.8.4.4} Mobile phones were instrumented to receive the DNS configuration from their operators; it follows that several operators (\eg Telenor in Pakistan, IND Airtel in India, and Telcel in Mexico) sometimes rely on Google DNS. Note that this event is quite rare (less than 1\% of the measurements) with the exception of Telkom (Kenya), where Google DNS was reported in 27\% of the measurements. 

Figure~\ref{fig:dns} shows that the median duration of DNS lookups with operator-provided DNS is quite fast, about 40ms. This is likely due to the high popularity of the websites under test (see Section~\ref{sec:data:meth}), which can be resolved from the DNS cache. When comparing mobile operators, their DNS achieve overall similar performance. Conversely, Google DNS lookups are much slower (up to a  10x increase for Telcel in Mexico). We conjecture that Google DNS is used as a backup solution in presence of outages or high load, since its usage was rare in our experiments. 


\vspace{0.1in}
\noindent\textbf{Content Delivery Networks}: We study the performance of each mobile operator towards six popular CDNs by downloading the last version of \texttt{jquery.min.js} hosted by each of them. On average, this file was downloaded about 117 times per CDN; all downloads were carried out over HTTP/2 with TLSv1.3, using various cipher suites.


We start by comparing the performance of each CDN globally, \ie aggregating measurements from \nummobile mobile networks (see Figure~\ref{fig:CDN_aggregate_hit_miss}). When possible (\texttt{Cloudflare}, \texttt{jsDelivr}, and \texttt{ Microsoft Ajax}), we differentiate between a cache HIT (solid box plots), and a cache MISS (hatched box plots), which are identified using two HTTP response header fields:  \texttt{x-cache} and \texttt{cf-cache-status} (for \texttt{Cloudflare}~\cite{cloudflare}). The lack of hatched box plot for \texttt{Microsoft Ajax} indicates absence of cache misses, while for the remainder CDNs no distinction was possible. 


If we focus on cache HIT, Figure~\ref{fig:CDN_aggregate_hit_miss} shows that the fastest download times are achieved via \texttt{jsDelivr}, which saves a minimum of 100~ms (when compared with \texttt{Cloudflare}) and up to 700~ms (when compared with \texttt{Highwinds}). Our results confirm that \texttt{jsDelivr} does indeed perform the best as they claim -- since it relies on a network of CDNs. If we focus on cache MISS, the download time increases significantly, about 3x for both \texttt{Cloudflare} and \texttt{jsDelivr}. Still, even in presence of misses both CDNs manage to be faster than \texttt{Highwinds}.




Next, we focus on cache HIT only and we analyze the CDN performance by mobile network. Given we identified clear patterns by continent, Figure~\ref{fig:CDN_aggregate_continent} shows one box plot for the total download time per continent: Europe, Australia, Asia, Central and South America, and Africa. The figure shows that, regardless of the CDN, the median download time grows by 10x (from 200-300ms up to 1-2 seconds) when comparing Europe and Africa. This result is the realization of networking effects we studied in the previous sections, such as high latencies and losses. For example the median latency (\texttt{mtr} to popular services as in Figure~\ref{fig:MTR_aggregate_ping}) for Ethio Telecom (Ethiopia), MTN Connect (Nigeria), and Telkom (Kenya) is 147ms, 108ms, and 149ms, versus 15ms, 23ms, and 50ms, in Play (Poland), Vodafone.de (Germany), and Lycamobile (France).

When comparing different CDNs, we confirm the high performance achieved by \texttt{jsDelivr}, which is only outperformed by \texttt{Google} and \texttt{Microsoft Ajax} in Africa. We can also see that \texttt{Highwinds} is competitive in Europe, Central and South America, and Africa, but suffers from very bad performance in Australia and Asia which skew its results when aggregating all locations (see Figure~\ref{fig:CDN_aggregate_hit_miss}). 

\begin{figure}[b]
    \centering
    \includegraphics[width=\linewidth]{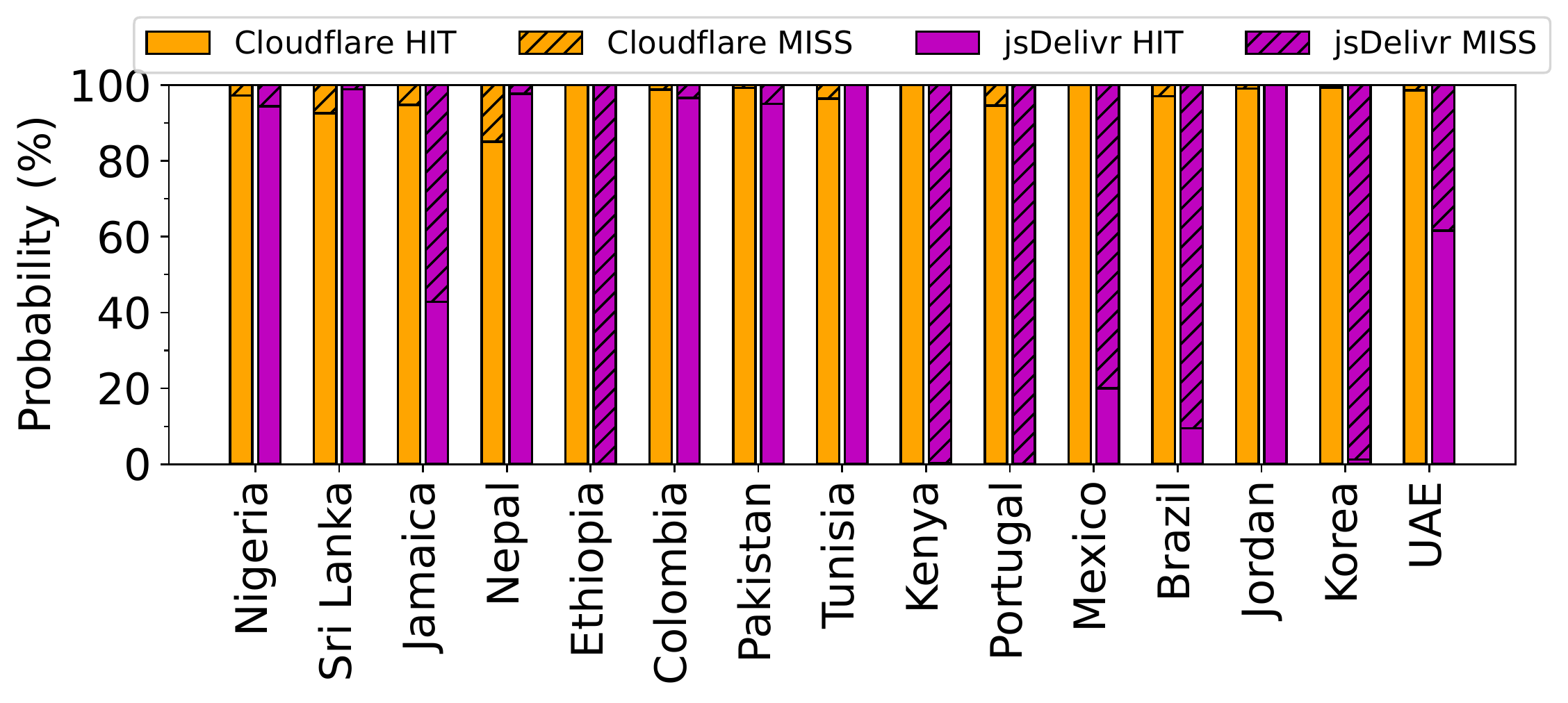}
    \caption{Probability of cache HITs versus MISSes across mobile networks for \texttt{Cloudflare} and \texttt{jsDelivr} CDNs. }\label{fig:hit_miss_split}
\end{figure}

Finally, Figure~\ref{fig:hit_miss_split} shows how the MISSes are distributed across both \texttt{Cloudflare} and \texttt{jsDelivr} for the mobile networks that experienced cache MISSes. The results shows that for \texttt{Cloudflare} there is a very small probability of having a cache MISS. In contrast, for \texttt{jsDelivr} it can be observed that a number of mobile networks (Ethio Telecom in Ethiopia, Telkom in Kenya, Vodafone in Portugal, and SK Telecom in Korea) have almost a 100\% cache MISSes for all tests performed at these locations. 



\begin{figure*}[t]
    \centering
    \includegraphics[width=\linewidth]{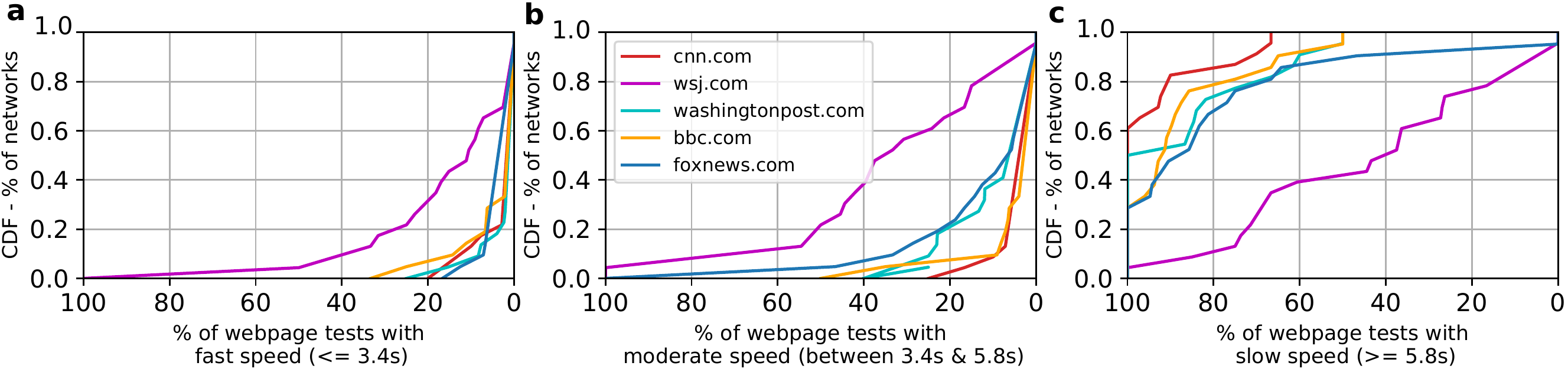}
    \caption{CDFs of percentage of mobile networks with web browsing speed index, across five popular news websites, classified as: (a) fast ($\le3.4s$), (b) moderate (between $3.4s$ and $5.8s$), (c) slow ($\ge5.8s$).}
    \label{fig:webpages_aggregate}
    \vspace{0.2in}
\end{figure*}

\section{Application Performance Analysis}
\label{sec:res:perf}
\vspace{0.05in}
\noindent\textbf{Web Browsing:} Figure~\ref{fig:webpages_aggregate} shows  Chrome UX style-based results for speed index~\cite{speedindex} split by tested news website (see Section~\ref{sec:data:meth} for more information). We use Google's lighthouse definition~\cite{lighthouse} of page speed: fast ($<=3.4s$), moderate (between $3.4s$ and $5.8s$), slow ($>= 5.8s$).

Figure~\ref{fig:webpages_aggregate}(a) shows that \textit{fast} webpage loads are rare: only 10\% of the mobile networks have 10\% of webpage loads faster than 3.4 seconds for four out of the five visited news web sites. The only exception is \texttt{wsj.com} which loads fast for 20\% of the measurements from 40\% of the mobile networks. Figure~\ref{fig:webpages_aggregate}(b) also shows a slight increase in the number of networks with a significant number of \textit{moderate} page loads: 20\% of the networks have nearly 55\%, 25\% and 20\% of their web tests considered to be of moderate speed for \texttt{wsj.com}, \texttt{foxnews.com}, and \texttt{washingtonpost.com}, respectively. Conversely, \texttt{bbc.com} and \texttt{cnn.com} still only have 10\% of fast page loads in 10\% of the mobile networks. Finally, 80\%-100\% of the webpages' tests are \textit{slow} for 70\% of the mobile networks (Figure~\ref{fig:webpages_aggregate}(c))--- except for \texttt{wsj.com}, again.  

The reason of such slow loading times is that news websites are quite \textit{heavy} for a low-end mobile device, \ie high CPU and memory usage. Not all news websites are the same, with  \texttt{wsj.com} largely outperforming the rest, and  \texttt{cnn.com} trailing, \ie it loads slow 90\% of the times for 90\% of the mobile networks. This is because \texttt{cnn.com} is one of the most complex pages with many embedded elements and recursive JavaScript. Given the importance of a diversified news outlet, such performance gap might have an impact on the user which goes far beyond their quality of experience.


\vspace{0.1in}
\noindent\textbf{Video Streaming:} We now focus on video streaming via YouTube. Overall, a \textit{good} streaming quality (480p--720p) is measured across most mobile networks. Lower quality levels are rare, with the exception of Lycamobile (France) and Claro (Colombia) where most streams only achieve 240p and 360p. We did not record any video stream at maximum resolution (1080p or 5~Mbps), despite many mobile operators offer download speeds which should support such quality. If we focus on the buffer playtime for mobile networks were high video qualities are detected, we observe median buffering time comprised between 30 and 60 seconds, which should be plenty to motivate the player to switch to a higher quality. We conjecture that the limited device resources have an impact on such switch, or some mobile operators adopt rate limits -- which is not unlikely, as discussed in~\cite{li2019large}. 

\begin{figure}[t]
    \centering
    \includegraphics[width=0.8\linewidth]{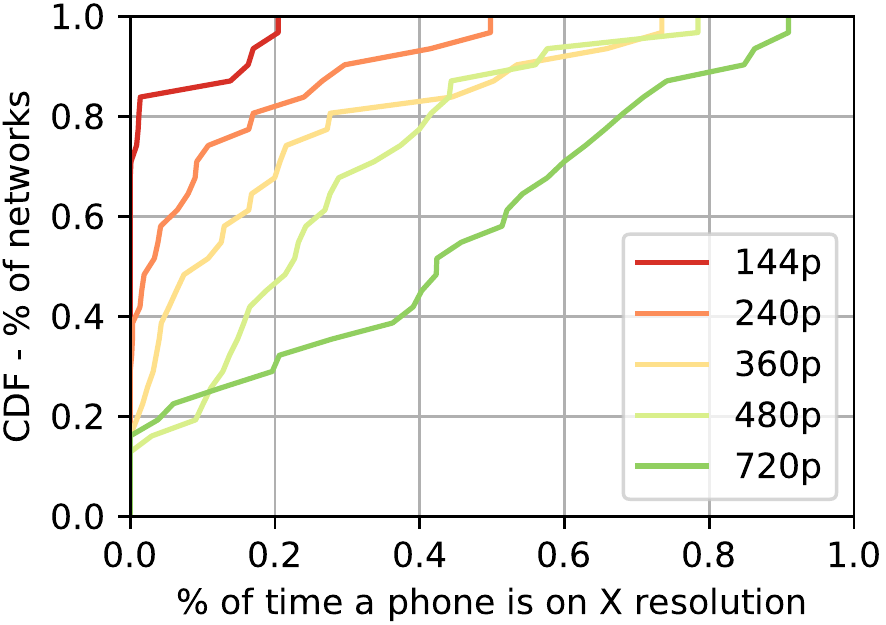}
    \caption{Percentage of mobile networks with a YouTube tests' probability distribution of being on  resolution comprised between 144p and 720p.}
    \label{fig:prob_youtube_res}
\end{figure}

We next generalize the above observations by analyzing the percentage of times that a stream quality is detected at each mobile operator (see Figure~\ref{fig:prob_youtube_res}). The figure shows that 1080p is indeed never detected, as discussed above, and that a \textit{poor} quality (144p) is rare: only 20\% of the mobile operators have up to 20\% of streams at this quality. Indeed, the probability of higher quality streams tend to increase with the quality, \eg half of the operators have at least 40\% of the streams at 720p. 

The above result suggests that YouTube performs much better across the different mobile networks in comparison to other previous conducted tests such web browsing. Given that YouTube quality does not highly depend on the network latency but rather on the available bandwidth (unlike web browsing). Ironically, we live in a world where it is much easier to watch a dancing cat\footnote{\url{https://www.youtube.com/watch?v=NUYvbT6vTPs}} rather than browsing a news article online.


\vspace{0.2in}
\section{Related Work}
\label{sec:related}
To the best of our knowledge, the closest related work to our study is~\cite{related_hung} by Hung et.\ al. This work studies which factors impact user perceived performance of smartphones' network applications (Web browsing, video streaming, and voice over IP). The study was conducted in 2010 over the 3G networks of four major U.S. mobile operators: AT\&T, Sprint, Verizon, and T-Mobile. In contrast, our paper studies and compares the performance of \numcountries operators across the globe. Further, we focus on 4G and introduce a broader set of networking experiments. Last but not least, we also propose a novel 
test-bed design based on regular Android devices. 


With respect to test-bed design and deployment, MONROE~\cite{ozgu2015monroe} is the closest approach to \testbed. MONROE is a measurement platform whose goal is to provide open-access assessment of the performance and reliability of mobile operators. MONROE is currently deployed across eleven mobile networks in four European countries. The core measurement component is called a \textit{node} which is equipped with a Debian-based single-board computer plus three LTE modems connected to different providers. A centralized scheduling system allows MONROE users to post custom-made experiments to distributed nodes and remotely collect measurement results. In addition, each node independently executes certain background experiments, such as RTT measurements, to MONROE servers.  Other large-scale research measurement platforms such as RIPE Atlas~\cite{staff2015ripe}, BISmark~\cite{sundaresan2014bismark}, PlanetLab~\cite{chun2003planetlab}, and GENI~\cite{berman2014geni} share common objectives with MONROE but are not designed for mobile environments. 



Finally, crowd-sourcing approaches like Netalyzr~\cite{kreibich2010netalyzr}, NetPiculet~\cite{wang2011untold}, OpenSignal~\cite{opensignal}, RootMetrics~\cite{rootmetrics} or MobiPerf~\cite{Mobiperf} use custom-designed apps for measuring mobile operators via crowdsourcing. \amigo uses a different approach since it allows full control of the device, \eg even execution of classic Linux applications via \texttt{termux}, with no user interference. Note that while the only role of \amigo's volunteers is to carry the device, provide connectivity opportunities, and keep it charged, they occasionally interferred with our experiments as discussed in the next section.

\vspace{0.15in}
\section{Conclusion and Lessons Learned}
\label{sec:conclusion}
The lack of public data or test-bed makes it difficult to compare the user experience from mobile operators around the world. This paper has proposed a solution to this problem by introducing a new test-bed design called \amigo. The design relies on travelers carrying mobile phones to act as measurement endpoints (MEs) and run a set of desired experiments. This design makes \amigo easy to deploy, while offering realistic user mobility and devices. All code behind \amigo~\cite{amigoCode}has been open sourced to help measurement researchers deploying their own test-beds in the wild. 

To demonstrate the \amigo capabilities, we have deployed it across five continents leveraging \numparticipants  students, and investigated the performance of \numcountries mobile networks. Among the many observations, we find that being \textit{consistently} fast is challenging for a mobile operator, both in term of download speed and access latency. It follows that applications requiring high bandwidth and low latency may suffer from frequent slowdowns. 

\amigo was also employed in~\cite{varvello2022performance} to examine the effectiveness of various videoconferencing applications across WiFi and mobile networks worldwide. The study discovered that mobile clients demonstrate comparable patterns to fixed-access clients concerning round-trip time  and video rate. Nevertheless, the overall quality of experience (QoE) at the application level is significantly affected by the limited processing power and screen size of mobile devices. \amigo was also used in~\cite{tags} to assess the performance of the two most popular location tags (Apple's AirTag and Samsung's SmartTag) through in-the-wild experiments conducted by having these tags as part of the vantage points carried by the amigos.

During the experiments conducted in this paper, we encountered a number of issues/challenges that we summarize below. 

\vspace{0.15in}
\noindent \textbf{Battery Charging is Crucial:}  Testing mobile networks can place a considerable strain on battery life, particularly for low-end devices such as the \redmi.  As participants carried MEs without frequently interacting with them, it was common for the devices to run out of power unnoticed. To address this issue, we devised a strategy that would halt experiments when the battery level dropped below 15\%, subsequently notifying participants through the installed app (see Figure~\ref{fig:amigo_testbed_gui}). This solution led to a 50\% reduction in the amount of time measurement endpoints were unreachable due to dead batteries. 

\vspace{0.15in}
\noindent \textbf{Debugging in the Wild:} Given the highly dynamic nature of the \testbed test-bed, we frequently encountered various unexpected issues. For example, ISP-specific policies or unanticipated participant behaviors (see below) that interfered with our measurements. To address these concerns, it was crucial to establish remote access to the MEs. Given all the devices in the field were situated behind ISP NATs, we leveraged a reverse SSH tunnel requested as an automation instruction (see Section~\ref{sec:sys:server}). 


\vspace{0.15in}
\noindent \textbf{Unexpected Participants Behaviors:}  \testbed's MEs do not require user intervention, apart from setting up WiFi or mobile connectivity. However,  we still had to deal with unexpected user interactions with the MEs. For example, while we ensured no sound is played by YouTube, some users opted to set their device in ``no-disturb'' mode which triggered a bug in our code when disabling audio. In turn, this caused some of our experiments to play sound causing distress (and complaints) in our volunteers. Similarly, some users turned off the screen of their device when they noticed an experiment was running, which could impact our results. We heavily rely on data processing and visual inspection of collected screenshots to ensure proper data sanitization. During the measurement campaign, we further sent reminders asking our volunteers to avoid interacting with the device, if possible.

\bibliographystyle{abbrv} 
\bibliography{biblio}


\end{document}